\documentclass{article}%
\usepackage{amsfonts}
\usepackage{graphicx}
\usepackage{amsmath}
\usepackage{amssymb}%
\usepackage{booktabs}
\usepackage{multirow}
\providecommand{\U}[1]{\protect\rule{.1in}{.1in}}
\newtheorem {theorem}{Theorem}[section]
\newtheorem {proposition}{Proposition}[section]
\newtheorem {corollary}{Corollary}[section]

\newtheorem{definition}{Definition}[section]

\newtheorem{remark}{Remark}[section]

\newcommand{\E}{\mathbb{E}}

\newcommand{\bi}[1]{\mbox{\boldmath{$ #1 $}}}

\begin{document}

\begin{center}

{\LARGE Gini Covariance Matrix and its Affine \\[1ex]
Equivariant Version}

\bigskip

\centerline{\today}

\bigskip Xin Dang$^{1,3}$\footnotetext[3]{Corresponding author}, Hailin Sang$^{1}$ and Lauren Weatherall$^{2}$

\bigskip $^1$Department of Mathematics, University of Mississippi,
University, MS 38677, USA. E-mail addresses: xdang@olemiss.edu, sang@olemiss.edu \\
$^2$ BlueCross \& BlueShield of Mississippi, 3545 Lakeland Drive, Flowood, MS 39232, USA.
E-mail address: laweatherall@bcbsms.com

\bigskip
\end{center}

\bigskip\textbf{Abbreviated Title: }{\Large Gini Covariance Matrix}

\begin{center}
\bigskip\textbf{Abstract}
\end{center}

We propose a new covariance matrix called Gini covariance matrix (GCM), which is a natural generalization of univariate Gini mean difference (GMD) to the multivariate case. The extension is based on the covariance representation of GMD by applying the multivariate spatial rank function.  We study properties of GCM, especially in the elliptical distribution family. In order to gain the affine equivariance property for GCM, we utilize the transformation-retransformation (TR) technique and obtain an affine equivariant version GCM that turns out to be a symmetrized M-functional. The influence function of those two GCM's are obtained and their estimation has been presented. Asymptotic results of estimators have been established. A closely related scatter Kotz functional and its estimator are also explored. Finally, asymptotical efficiency and finite sample efficiency of the TR version GCM are compared with those of sample covariance matrix, Tyler-M estimator and other scatter estimators under different distributions. 

\vspace{0.35cm}

\vspace{0.35cm}


\section{Introduction}
Gini mean difference (GMD) was introduced by Corrado Gini in 1914 as an alternative measure of variability. Since then, GMD and its derivatives such as Gini index have been widely used in a variety of research fields especially in finance, economics and social welfare (Yitzhaki and Schechtman, 2013). Rather than the assumption on the finite second moment, the GMD only requires existence of the finite mean of the distribution (Yitzhaki, 2003).  Hence GMD is more robust than the variance and it is often used for heavy-tailed asymmetric distributions, although it is less robust than some scale measures without any moment conditions. On the other hand, GMD is highly efficient. The relative efficiencies (RE) of the sample GMD with respect to sample standard deviation are about 0.98 under the normal distributions, 1.21 under the Laplace distribution and 1.86 under the $t(5)$ distribution (Nair, 1936; Gerstenberger and Vogel, 2015). With a little loss on efficiency, GMD gains robustness against departures from normal distributions.

In this paper, we extend GMD to the multivariate case. We propose the Gini covariance matrix (GCM) as (2 times) the covariance of $\bi X$ with its spatial rank $\bi r(\bi X)$, which is a direct generalization from a covariance representation of the univariate GMD. While the covariance matrix (Cov) is the covariance of $\bi X$ with itself and the rank covariance matrix (RCM) (Visuri {\em et al.}, 2000) is the covariance of $\bi r(\bi X)$ with $\bi r(\bi X)$, intuitively GCM is a new scatter measure between Cov and RCM.  With no surprise, the efficiency and robustness of sample GCM are between those of sample Cov and RCM. In terms of balance between efficiency and robustness, sample Gini covariance matrix provides us an extra method for multivariate statistical inference including multivariate analysis of variance, principle component analysis, factor analysis, and canonical correlation analysis.

As any estimator based on spatial signs and ranks, GCM is only orthogonally equivariant. In order to gain fully affine equivariant property,  we utilize a transformation-retransformation (TR) technique (Charkraborty and Chaudhuri, 1996;  Serfling, 2010) to obtain an affine equivariant version of GCM.  The well-known scatter Tyler M-functional (Tyler, 1987) is a TR version of the spatial sign covariance matrix. D\"{u}mbgen (1998) considered symmetrized TR spatial sign covariance matrix on the difference of two independent vectors $\bi X_1-\bi X_2$. D\"{u}mbgen {\em et al.} (2015) provided a general treatment on M-functionals of scatter based on symmetrizations of arbitrary order.  Our TR Gini covariance matrix turns out to be a pairwise symmetrized scatter M-functional. Compared to the regular M-functional, the symmetrized one has several advantages as emphasized in Sirki\"{a} {\em et al.} (2007). The distribution of pairwise differences is symmetric at $\bi 0$, hence avoids imposing some arbitrary definition of location for non-symmetric distributions. For elliptical distributions, there is no need to estimate location simultaneously for scatter M-estimators. 
Hence they avoid restrictive regularity conditions for joint existence of location and scatter estimators and may take fewer iterations to converge than their counterparts. Further,  a symmetrized scatter matrix has the so-called block independence property:  it is a block diagonal matrix if the block components of the random vector are independent.  Such a property holds naturally for the regular covariance matrix but may not for general M-functionals and some robust alternatives (Nordhausen and Tyler, 2015). Both versions of GCM have the block independence property and hence can be applied to independent component analysis (Hyv\"{a}rinen {\em et al.},  2001) or invariant coordinate selection (Tyler {\em et al.}, 2009). The price to pay for those advantages of the pairwise difference approach is an increase of computation burden and a loss of some robustness. If a procedure has computation complexity $O(n)$, its symmetrized one may require $O(n^2)$,  although D\"{u}mbgen {\em et al.}  (2016) have presented new algorithms for symmetrized M-estimators to reduce the computation time substantially. For large $n$, they approximate symmetrized estimators by considering the surrogate ones rather than all pairwise differences. Their algorithms can easily be adopted for our estimators. The decrease of robustness in symmetrized procedures seems to be understandable since one single outlier affects $n-1$ pairwise differences.  One must take into consideration of efficiency, robustness and computation when choosing a proper procedure for applications at hand. 

Koshevoy {\em et al.}  (2003) considered other multivariate extensions of mean deviation and Gini mean difference using the geometric volumes of zonotopes and lift-zonotopes. Their covariance matrices share many similar properties as our proposed ones, but ours enjoy simplicity and computational ease.  Although the approach differs from ours, it is worthwhile to mention that Serfling and Xiao (2007) also generalize GMD to multivariate case through L-moment approach. 

The remainder of the paper is organized as follows. In Section 2, we first review the Gini mean difference and the spatial rank function, then introduce the Gini covariance matrix and its affine equivariant version. Section 3 explores the influence functions of the two Gini covariance matrices. Section 4 presents estimation of the two Gini covariance matrices and asymptotical properties of the estimators. Asymptotic and finite sample efficiencies of the proposed TR Gini covariance estimator have been studied and compared with other estimators. The paper ends with some final comments in Section 5. All proofs are reserved to Appendix Section.

\section{Two Gini Covariance Matrices}
\subsection{Gini Mean Difference}
Gini mean difference (GMD) was introduced as an alternative measure of variability to the usual standard deviation. For a random variable $X$ from a univariate distribution $F$,  the GMD of $X$ (or $F$) is
\begin{equation}\label{eqn:gmd}
\sigma_g=\sigma_g(X)=\sigma_g(F)=\E |X_1-X_2|,
\end{equation}
where $X_1$ and $X_2$ are independent random variables from $F$. In contrast, the variance of $X$ (or $F$) is
$$\sigma^2_v(F)=\mbox{var}(X)= \frac{1}{2}\E (X_1-X_2)^2.$$
Rather than the assumption of finite second moment, the GMD only requires existence of a finite mean of $F$.  Hence the Gini mean difference is often used for heavy-tailed asymmetric distributions, especially in social welfare and the fields of decision-making under financial risk.

Among many representations such as Lorenz curve or $L$-functional formulations (Yitzhaki and Schechtman, 2013), we are interested in covariance formulations. One of them is
\begin{equation*}\label{eqn:gmdf}
\sigma_g(F)=4\mbox{Cov}(X, F(X)).
\end{equation*}
While the variance is the covariance of $X$ with itself, the GMD is (4 times) the covariance of $X$ with $F(X)$. In this spirit,  two Gini-type alternatives to the usual covariance for measuring the dependence of random variable $X$ and another random variable $Y$ with distribution function $H$  are
\begin{equation*}
\mbox{Cov}_g(X,Y)=4\mbox{Cov}(X, H(Y)), \;\; \mbox{Cov}_g(Y,X)=4\mbox{Cov}(Y, F(X)).
\end{equation*}
Such extensions are natural and useful (Yitzhaki, 2003; Carcea and Serfling, 2015). However, a major drawback is the asymmetry between $X$ and $Y$, i.e., $\mbox{Cov}_g(X,Y) \neq\mbox{Cov}_g(Y,X)$, in general. An even worse part is that $\mbox{Cov}_g(X,Y)$ and $\mbox{Cov}_g(X,Y)$ may have different signs in some cases (Yitzhaki, 2003), which brings substantial difficulty in interpretation. The asymmetry stems from the usage of $F(X)$ or $H(Y)$, which can be thought as a standardized marginal rank. A symmetry one calls for a `joint' rank of $X$ and $Y$.  

The other covariance type formulation for GMD is
\begin{equation*}\label{eqn:gmd2f}
\sigma_g(F)=2\mbox{Cov}(X, 2F(X)-1),
\end{equation*}
allowing an insightful interpretation: $\sigma_g(X)$ is twice of the covariance of $X$ and the {\em centered} rank function $r(X)=2F(X)-1$. $r(X)$ is centered because $\E r(X)=0$ if $F$ is continuous. So
\begin{equation}\label{eqn:gmdr}
\sigma_g(F)=2\mbox{Cov}(X, r(X))=2 \E (Xr(X)).
\end{equation}
A nice generalization of the centered rank in high dimensions provides a joint rank, and along with the representation of GMD in \eqref{eqn:gmdr} yields a natural extension of GMD for a multivariate distribution $F$.  

\subsection{Spatial Rank Function} \label{sec:sr}

Let $\bi X$ be a $d$-variate random vector from a continuous distribution $F$ with a finite first moment and the expected Euclidean distance from $\bi x$ to $\bi X$ be $D(\bi x, F)=\E_F\|\bi x-\bi X\|$. Then the gradient of $D$ is denoted as the centered {\em spatial rank function} (M\"{o}tt\"{o}nen {\em et al.}, 1997), that is,
 \begin{equation}\label{eqn:sr}
\bi r(\bi x)=\nabla_{\bi x} D(\bi x, F)=\E \frac{\bi x-\bi {X}}{\|\bi x-\bi X\|}=\E \{\bi s(\bi{x}-\bi{X})\},
\end{equation}
where $\bi s(\bi{x})=\bi{x}/\|\bi{x}\|$ $(\bi s(\bi{0})=\bi{0})$ is the
{\em spatial sign function} in $\mathbb{R}^d$. The spatial rank function is  the {\em expected direction} from $\bi X$ to $\bi x$. We call it centered because a random rank is centered at $\bi 0$, that is, $\E \bi r(\bi X)=\bi 0$. The solution of $\bi x$ in $\bi r(\bi x)=\bi 0$ is called the spatial median of $F$, which minimizes $D$. In the univariate case, the derivative of $D(x,F)=\E |x-X|$ with respect to $x$ leads to the univariate centered rank function $r(x)=\E \mbox{sign}(x-X)=2F(x)-1\in [-1,1]$ if $F$ is continuous. Clearly, the median of $F$ has a center rank 0.

The spatial rank function has many nice properties. The rank function $\bi r(\bi x)$ characterizes the distribution $F$ (up to a location shift) (Koltchinskii, 1997;  Oja, 2010), which means that if we know the rank function, we know the distribution (up to a location shift). Under weak assumptions on $F$, $\bi r(\bi x)$ is a one-to-one mapping from $\bi x$ to a vector inside the unit ball with the magnitude $\|\bi r(\bi x)\| \in [0,1]$ and the center of the unit ball is the spatial median of $F$.

Marginal ranks are less interesting because they are neither rotation nor scale equivariant. Also they lack the efficiency at the normal model since they loss dependence information (Visuri {\em et al.}, 2000). Based on the expected geometric volume of the simplex formed by $\bi x$ and $d$ random vectors from $F$, i.e, $D(\bi x, F)=\E_F V(\bi x, \bi X_1, ..., \bi X_d)$, Oja rank and sign functions are defined analogously (Oja, 1983). The major concern of this joint rank function is the computation of its sample version, especially for high dimensions.  However, the sample spatial rank is simple and easy to compute, which makes it advantageous and feasible in practice. For this reason, we will use the spatial rank function to define our multivariate Gini covariance matrix.


\subsection{Gini Covariance Matrix}
\begin{definition}
For a $d$-variate random vector $\bi X$ from a continuous distribution $F$ possessing a finite first moment, the Gini covariance matrix of $\bi X$ or $F$  (GCM) is defined as
\begin{equation*}
\bi \Sigma_g=\bi\Sigma_g(F)=2\E[ \bi X\bi r^T(\bi X)], 
\end{equation*}
where $\bi r(\bi x)$ is the spatial rank function defined in (\ref{eqn:sr}). 
\end{definition}
 As we can see, the definition of the Gini covariance matrix is a direct generalization from \eqref{eqn:gmdr}. Equivalently, let $\bi X_1$ and $\bi X_2$ be independent random vectors from $F$ and $\bi s(\bi x)$ be the spatial sign function, then we have 
\begin{align} 
&\bi \Sigma_g=2\E\bi X_1\E [\bi s^T(\bi X_1-\bi X_2)|\bi X_1]=2\E[ \bi X_1\bi s^T(\bi X_1-\bi X_2)] \nonumber\\
&=\E\frac{(\bi X_1-\bi X_2)(\bi X_1-\bi X_2)^T}{\|\bi X_1-\bi X_2\|}. \label{eqn:gcm}
\end{align}
The third equality in \eqref{eqn:gcm} is a result of
$$\E\frac{\bi X_1(\bi X_1-\bi X_2)^T}{\|\bi X_1-\bi X_2\|}=-\E\frac{\bi X_2(\bi X_1-\bi X_2)^T}{\|\bi X_1-\bi X_2\|}.$$
From \eqref{eqn:gcm}, it is easy to prove that $\bi\Sigma_g$ is positive definite since $F$ is continuous. 
Equation \eqref{eqn:gcm} additionally recovers the $L_1$ metric representation of Gini mean difference \eqref{eqn:gmd} when $d=1$. It also demonstrates that it is a pairwise difference approach defined without reference to a location parameter.

From \eqref{eqn:gcm}, we can also write $\bi\Sigma_g$ as $\E(\bi X_1-\bi X_2)\bi s^T(\bi X_1-\bi X_2)$, which is the expected matrix of the product of a pairwise difference and its directional sign function.  If we only use directional information of $F$,  we obtain
$$
\E \bi s(\bi X_1-\bi X_2)\bi s^T(\bi X_1-\bi X_2)=\E\frac{(\bi X_1-\bi X_2)(\bi X_1-\bi X_2)^T}{\|\bi X_1-\bi X_2\|^2}.
$$
The resulting matrix is known as the symmetrized spatial sign covariance matrix (SSCM), which has been studied by Visuri {\em et al.} (2000), Croux {\em et al.} (2002) and Taskinen {\em et al.} (2012). 

The spatial rank covariance matrix (RCM) is defined as the covariance matrix of spatial rank. That is,
\begin{align}\label{eqn:rcm}
\E\bi r(\bi X)\bi r ^T(\bi X)=\E \bi s(\bi X_1-\bi X_2)\bi s^T(\bi X_1-\bi X_3)=\E\frac{(\bi X_1-\bi X_2)(\bi X_1-\bi X_3)^T}{\|\bi X_1-\bi X_2\|\|\bi X_1-\bi X_3\|}.
\end{align}
RCM and its modified version have been studied by Visuri {\em et al.} (2000) and Yu {\em et al.} (2015).  Since RCM uses three independent random vectors in its definition, the sample RCM is more efficient than the sample SSCM. 


Before we explore properties of the Gini covariance matrix, it is worthwhile to note that another useful extension of GMD from the covariance representation based on the spatial rank function is $2\E\bi X^T\bi r(\bi X)$. This generalization coincides with the multivariate Gini mean difference defined in Koshevoy and Mosler (1997). That is,
$2\E\bi X^T\bi r(\bi X)=\E \|\bi X_1-\bi X_2\|$.  Clearly, it is also an immediate extension from the $L_1$ metric representation of \eqref{eqn:gmd}.  


\subsection{Properties of Gini covariance matrix}

We study properties of the Gini covariance matrix under elliptical distributions.
\begin{definition}
 A $d$-variate absolutely continuous random vector $\bi X$ has an elliptical distribution if its density function is of the form
\begin{equation}
f(\bi x| \bi \mu, \bi\Sigma)
 =|\bi\Sigma|^{-1/2} g\{(\bi x-\bi {\mu})^T\bi\Sigma^{-1}(\bi x-\bi{\mu})\},\label{eqn:elliptical}
\end{equation}
for some positive definite symmetric matrix $\bi \Sigma$ and nonnegative function $g$ with $\int_0^\infty t^{d/2-1}g(t)dt < \infty$. 
\end{definition}
If $t^{(d-k)/2-1}g(t)$ is integrable, then the $k^{th}$ moment of $\bi X$ exits. The parameter $\bi{\mu}$ is the symmetric center and it equals the first moment if it exists. The scatter parameter $\bi\Sigma$ is proportional to the covariance matrix when it exists. It should be noted that elliptical distributions can also be defined through the characteristic functions without assuming densities. 
In addition, the variates $R=\|\bi \Sigma^{-1/2}(\bi X-\bi \mu)\|$ and $\bi U=\{\bi \Sigma^{-1/2}(\bi X-\bi \mu)\}/R$ are independent with $\bi U$ being uniformly distributed on the unit sphere and  $R$ having density
\begin{equation}\label{eqn:pdfr}
 f_r(r)=\frac{2\pi ^{d/2}}{\Gamma(d/2)} r^{d-1} g(r^2),
 \end{equation}
where $\Gamma(a)=\int_0^\infty t^{a-1}e^{-t}dt$ is the gamma function. The independence of $R$ and $\bi U$ follows from Lemma  1 of Stamatis {\em et al.} (1981). Note that if the covariance matrix of $\bi X$ exists, it equals $(\E R^2/d) \bi \Sigma$.  More details on the elliptical distribution family refer to Fang and  Anderson (1990).

The family of elliptical distributions is denoted as ${\cal E}(\bi\mu, \bi\Sigma, g)$. If $\bi \mu=\bi 0$ and $\bi \Sigma=\bi I_{d}$ (the $d\times d$ identity matrix), we call the distribution spherically symmetric and denote it as $F_0(g)$.

The family of elliptical distributions contains a quite rich collection of models. Perhaps the most widely used one is the Gaussian distribution, in which
\begin{equation*}\label{eqn:normal}
g(t)=(2\pi)^{-d/2}e^{-t/2}.
\end{equation*}
Other than that, $t$ distributions are commonly used in modeling data with heavy-tailed regions. In the case of the $t$ distributions,
$$g(t)=\frac{\Gamma[(\nu+d)/2]}{\Gamma(\nu/2)(\nu\pi)^{d/2}}(1+t/\nu)^{-(d+\nu)/2},$$
where $\nu$ is the degree of freedom parameter.  $\nu$ determines the fatness of the tail regions. For $\nu=1$, it is called $d$-variate Cauchy distribution,  which has very heavy tails where even the first moment does not exist. When $\nu \rightarrow \infty$, it yields the Gaussian distribution.

A quite flexible elliptical family is called Kotz type distributions (Kotz, 1975; Nadarajah, 2003), in which the density is of the form \eqref{eqn:elliptical} with
\begin{equation*}
g(t)=c(d,\alpha, \beta, \gamma) t^{\alpha-1}e^{-\gamma t^{\beta}}. \label{eqn:kotztype}
\end{equation*}
The parameters are $\beta, \gamma >0$, $\alpha>1-d/2$ and $c(d,\alpha, \beta, \gamma)$ is the normalization constant. Clearly, when $\beta=1$,  $\alpha=1$ and $\gamma=1/2$, the distribution reduces to the Gaussian distribution. The heaviness (or lightness) of tail regions of distributions mainly depends on $\beta$.
In particular, we take the special case of $\beta=1/2$, $\alpha=1$ and $\gamma=1$ for demonstration in later sections, that is,
\begin{equation}
g(t)=\frac{\Gamma(d/2)}{2 \pi^{d/2}\Gamma(d)}e^{-\sqrt{t}}. \label{eqn:kotz}
\end{equation}
We call it the Kotz distribution. For $d=1$, the Kotz distribution reduces to the Laplace distribution. It can be viewed as a multivariate generalization of Laplace distribution. Arslan (2010) also considered this distribution and extended it to asymmetry distributions by introducing a skewness parameter. 

The following theorem states the relationship of the Gini covariance matrix and the scatter matrix $\bi \Sigma$ in elliptical distributions.

\begin{theorem}\label{thm:eigv}
If $\bi X$ is elliptically distributed from $F$ with the first moment and the scatter parameter $\bi \Sigma$ having the spectral decomposition $V \Lambda V^T$, then $\bi \Sigma_g=V\Lambda_gV^T$ with
$$\Lambda_g=\mbox{diag}(\lambda_{g,1},...,\lambda_{g,d})=c(F) \E\left[\frac{\Lambda^{1/2}\bi U\bi U^T\Lambda^{1/2}}{\sqrt{\bi U^T\Lambda \bi U}}\right], $$
and hence $$ \lambda_{g,i}=c(F)\E\left[\frac{\lambda_i u_i^2}{\sqrt{\sum_{j=1}^d\lambda_j u_j^2}}\right],$$
where $\bi U=(U_1,...,U_d)^T$ is uniformly distributed on the unit sphere, $\lambda_i$'s are eigenvalues of $\bi \Sigma$ and $c(F)$ is a constant depending on distribution $F$.
\end{theorem}

The proof of Theorem \ref{thm:eigv} goes along the lines of the proof of Theorem 1 in Taskinen {\em et al.} (2012) and is given (together with all other proofs) in the Appendix. The main consequence is that the same orthogonal matrix $V$ diagonalizes both $\bi \Sigma$ and $\bi \Sigma_g$. In other words, the Gini covariance matrix $\bi \Sigma_g$ has the same eigenvectors as $\bi \Sigma$. Consequently, the Gini covariance matrix can be used for principal component analysis.

\begin{remark}\label{rem:identity}
In the case of an elliptical distribution $F$ in $ {\cal E}(\bi \mu,\bi \Sigma,g)$ having $\bi \Sigma = \bi I_{d}$,  it holds that $\lambda_{g,i}=c(F)\E (U_i^2/\| \bi U \|)=c(F)\E U_i^2=c(F)/d$ for all $i=1,...,d$ and hence $\bi \Sigma_g=
\frac{ c(F)}{d}\bi I_{d}$. In other words, for spherical distributions $F_0(g)$ (even $\bi \mu \neq \bi 0$), their Gini covariance matrix is the identity matrix multiplied by a factor. Dividing by this factor makes GCM estimator Fisher consistent to the scatter parameter at $F_0(g)$. An estimator $T(F_n)$ is Fisher consistent to $\theta$ if $T(\lim_{n\rightarrow \infty} F_n) =\theta$ where $F_n$ is the empirical distribution of sample $\bi X_1,...,\bi X_n$ from $F$.  
\end{remark}.

\begin{remark}\label{rem:cf}
For any elliptical distribution $F$ in ${\cal E}(\bi \mu, \bi \Sigma, g)$ and the associated spherical distribution $F_0(g)$,  the constant $c(F) =c(F_0)= \E_{F_0} \|\bi X_1-\bi X_2\|$, where $\bi X_1$ and $\bi X_2$ are independent random vectors from $F_0(g)$. Let $c_1(F_0) =\E_{F_0} \|\bi X\|$.  For Gaussian distributions, $c(F_0)=\sqrt{2}c_1(F_0)$. However, such a relationship may not hold for other elliptical distributions.  
\end{remark}

\begin{remark}\label{rem:normal}
If $F$ is a multivariate normal distribution ${\cal N}_d(\bi \mu,\bi \Sigma)$,
$c(F)=\sqrt{2}\E (D^{1/2}),$ where $D=(\bi X-\bi \mu)^T\bi\Sigma^{-1}(\bi X-\bi \mu)$ has a $\chi^2$ distribution with $d$ degrees of freedom. Hence
$c(F)={2\Gamma[(d+1)/2]}/{\Gamma(d/2)}.$
 For a univariate normal distribution ${\cal N}(\mu,\sigma^2)$, the Gini covariance is reduced to the Gini mean difference that equals $2\sigma/\sqrt{\pi}$.
\end{remark}

Spatial signs and spatial ranks are orthogonally
equivariant in the sense that for any $d\times d$ orthogonal  matrix $O$ ($O^T=O^{-1}$), $d$-dimensional vector $\bi b$ and nonzero scalar $c$,
letting $\bi X^*=c O\bi X+\bi b$ with the distribution $F^*$,
$$ \bi s(\bi X^*)=\mbox{sign}(c)O \bi s(\bi X), \;\mbox{ and } \;\;
  \bi r(\bi X^* , F^*)=\mbox{sign}(c)O\bi r(\bi X, F).$$
Therefore, we have the orthogonal equivariance property of GCM as follows.
\begin{align} \label{eqn:Ortho}
\bi \Sigma_g(cO\bi X+\bi b) =\bi \Sigma_g(F^*)= |c|O\bi \Sigma_g(\bi X)O^T.
\end{align}

Orthogonal equivariance property of the Gini  covariance matrix $\bi \Sigma_g$ holds under any distribution with a finite first moment. Orthogonal equivariance ensures that under rotation, translation and homogeneous scale change, the quantities are transformed accordingly. However, it does not allow heterogeneous scale changes. The equality does not hold for a general $d\times d$ nonsingular matrix $A$. Hence the Gini covariance matrix is not fully affine equivariant.

\subsection{The Affine Equivariant Version of GCM}

In order to achieve full affine equivariance, we use the transformation - retransformation (TR) technique,  which serves as  standardization of multivariate data. More details can be found in Charkraborty and Chaudhuri (1996) and Serfling (2010).  The affine equivariant counterpart of the Gini covariance matrix is denoted as $\bi\Sigma_G$. The idea is that if $\bi X_1$ and $\bi X_2$ are independent random vectors from $F$ and  they are transformed or standardized to be $\bi Z_i=\bi \Sigma_G^{-1/2}(\bi X_i-\bi\mu)$  for $i=1,2$, then $\bi Z_1$ and $\bi Z_2$ are independently distributed from the spherical distribution $F_0$ with the scatter matrix $\bi I_{d}$.  By Remark \ref{rem:identity}, we thus have
\begin{align}
\bi \Sigma_g(F_0)=\E\frac{(\bi Z_1-\bi Z_2)(\bi Z_1-\bi Z_2)^T}{\|\bi Z_1-\bi Z_2\|} =\frac{c(F)}{d} \bi I_d. \label{eqn:trgcm1}
\end{align}
Since $\bi Z_1-\bi Z_2=\bi \Sigma_G^{-1/2}(\bi X_1-\bi X_2)$, the middle term of (\ref{eqn:trgcm1}) is $$\E \frac{\bi \Sigma_G^{-1/2}(\bi X_1-\bi X_2)(\bi X_1-\bi X_2)^T\bi \Sigma_G^{-1/2}}{\sqrt{(\bi X_1-\bi X_2)^T\bi \Sigma_G^{-1}(\bi X_1-\bi X_2)}}. $$ 
Thus, the TR version of the GCM is defined as follows.  
\begin{definition}\label{def:afgcm}
For a $d$-variate elliptical distribution $F$ with existing first moment, its TR version of the Gini covariance matrix, denoted as $\bi \Sigma_G$,  is defined as the solution of
\begin{equation}\label{eqn:trgcm2}
\bi \Sigma_G= \frac{d}{c(F)}\E \frac{(\bi X_1-\bi X_2)(\bi X_1-\bi X_2)^T}{\sqrt{(\bi X_1-\bi X_2)^T\bi \Sigma_G^{-1}(\bi X_1-\bi X_2)}},
\end{equation}
where $\bi X_1, \bi X_2 \stackrel{iid}{ \sim} F$ and $c(F)$ is given in Remark \ref{rem:cf}. 
\end{definition}

\begin{theorem}\label{thm:afeq}
The matrix valued functional $\bi \Sigma_G(\cdot)$ is a scatter matrix in the sense that for any nonsingular $d\times d$ matrix $A$ and $d$-vector $\bi b$, $\bi \Sigma_G(F^*)=A\bi \Sigma_G(F)A^T$, where $\bi X$ is distributed from $F$ and $F^*$ is the distribution of $A\bi X+\bi b$. 
\end{theorem}

\begin{remark}
For $d=1$, the affine equivariant Gini mean difference $\sigma_G$ satisfies $\sigma_G^{1/2}= \E_F|X_1-X_2|/c(F)=\sigma$. Hence, $\sigma_G$ is Fisher consistent to the squared scale parameter for distributions in the location-scale family. Also $\sigma_G$ only assumes the existence of first moment compared to the second moment needed for the variance $\sigma_v^2=\E(X-\mu)^2=1/2\E(X_1-X_2)^2$.
\end{remark}

Note that the affine equivariant version of GCM $\bi \Sigma_G$ is a symmetrized M-functional.  Sirki\"{a} {\em et al.} (2007) studied a general symmetrized M-functional $\bi \Sigma_M$ that solves
$$ \E [w_1(R_{12}(\bi \Sigma_M)) \bi U_{12}(\bi \Sigma_M)\bi U_{12}^T(\bi \Sigma_M)-w_2(R_{12}(\bi \Sigma_M)) \bi I_d] = \bi 0,$$
where $w_1$ and $w_2$ are real-valued functions on $[0,\infty)$, $\bi X_1, \bi X_2\stackrel{iid}{ \sim} F$, $\bi Z_{12}(\bi \Sigma_M)=\bi \Sigma_M^{-1/2}(\bi X_1-\bi X_2)$, $R_{12}(\bi \Sigma_M)=\|\bi Z_{12}(\bi \Sigma_M)\|$ and $\bi U_{12}(\bi \Sigma_M)=R_{12}(\bi \Sigma_M)^{-1}\bi Z_{12}(\bi \Sigma_M)$.   Clearly,  the weight functions for $\bi \Sigma_G$ are $w_1(t)=t$ and $w_2(t)= c(F)/d$.  For the case of $w_1(t)=t^2$ and $w_2(t)=2$, the covariance matrix is obtained. If $w_1(t)=d$, $w_2(t) =1$ and an additional condition that the trace of the matrix is $d$,  the symmetrized Tyler M-functional called D{\" u}mbgen's M-functional is obtained (D{\" u}mbgen, 1998). D{\" u}mbgen's M-functional is also a TR-version of symmetrized spatial sign matrix. The very recent paper (D{\" u}mbgen {\em et al}., 2015) considers a framework to generalize M-functionals  based on symmmetrizations of arbitrary order. Our TR Gini covariance matrix can be treated as an example of their Case 1 with the symmetrization of order 2.  

As defined using pairwise differences, the symmetrized M-functional is obtained without reference to the location parameter $\bi \mu$. Maronna (1976) considered simultaneous location and scatter M-functionals. In this paper, since we focus on symmetrized scatter M-functionals, we assume $\bi \mu$ known for scatter M-functionals.  Letting $\bi Z(M)= M^{-1/2}(\bi X-\bi \mu)$, $R( M)=\|\bi Z(M)\|$ and $\bi U(M)=R(M)^{-1}\bi Z(M)$, we obtain a regular scatter M-functional that solves
$$ \E [w_1(R(M)) \bi U(M)\bi U^T(M)-w_2(R(M)) \bi I_d] = \bi 0.$$
For $w_1(t)=t^2, w_2(t)=1$, the covariance matrix is obtained. Tyler M-functional is the case of $w_1(t)=d$, $w_2(t) =1$ with an additional condition that the trace of the matrix $M$ is $d$. The case of $w_1(t)=t$ and $w_2(t)=1$ is called Kotz functional, denoted as $\bi \Sigma_K$.  The rational for such a name is because it equals the scatter parameter of the Kotz distribution (\ref{eqn:kotz}).  

Note that a $L_1$-type M-functional considered by Roelant and Van Aelst (2007) and Arslan (2010) is the simultaneous location and scatter Kotz M-functional. Our TR Gini covariance matrix can be viewed as the symmetrized Kotz functional. In other words, TR Gini covariance is a multivariate extension of $\E|X_1-X_2|$, while $\bi \Sigma_K$ is for $\E |X-\mu|$. It is worth to mention that Koshevoy {\em et al.} (2003) considered extensions of mean deviation and mean difference using volumes of zonotope and lift-zonotope. For $F\in {\cal E}(\bi \mu,\bi \Sigma, g) $, we have that $\bi \Sigma_K(F) =\{\E R /d\}^2 \bi \Sigma$ with  $R=\|\bi \Sigma^{-1/2}(\bi X-\bi \mu)\|$ and the zonoid covariance matrix (ZCM) $\bi \Sigma_Z(F)$ is equal to $c(d) \bi \Sigma_K(F)$ with the factor $c(d)$ depending on the dimension $d$ but independent on the distribution $F$. 

Instead of taking the TR technique, Visuri {\em et al.} (2000) re-estimated each eigenvalue of the spatial rank covariance (RCM) defined in \eqref{eqn:rcm} to make it affine equivariant. Yu {\em et al.} (2015) used the median of absolute deviation (MAD) to estimate scale of each univariate projected data on each of eigenvector directions. Thus the resulting RCM is affine equivariant and robust.  However, it may trade off too much efficiency for robustness.  The simulation in later section confirms its relatively low finite sample efficiency comparing to symmetrized M-estimators. Also, as cautioned in Nordhausen and Tyler (2015), those robust alternatives may 
not have the block independence property. 


\subsection{Block Independence Property}
One important property of symmetrized scatter functionals is the independence property, or more generally, the block independence property.  A scatter functional with the block independence property means that it is a block diagonal matrix if the block components of the random vector are independent.  Such a property holds naturally for the regular covariance matrix, but it may not hold for general M-functionals and some other robust scatter functionals, as noted in Nordhausen and Tyler (2015).  They proved that any symmetrized scatter functionals have the block independent property. In fact, such a result holds for any symmetrized orthogonally equivariant covariance matrix since the proof of their theorem only uses the conditions of symmetry and orthogonal equivariance.  As a result, our two versions of GCM have the block independence property as stated in the following corollary.

\begin{corollary}\label{thm:IP}
Let $\bi X^T =(\bi X_1^T, \bi X_2^T, ...,\bi X_k^T)$ have $k$ independent blocks with dimensions $d_1,...,d_k$ ($d_1+d_2+...+d_k=d$). Then $\bi \Sigma_g(\bi X)$ and $\bi \Sigma_G(\bi X)$ are block diagonal matrices with block dimensions $d_1,...,d_k$.
\end{corollary}

The block independence property is beneficial in many applications, for example, in independent component analysis (Oja {\em et al.}, 2006), in independent subspace analysis (Nordhausen and Oja, 2011), or in invariant coordinate selection (Tyler {\em et al.}, 2009). 

In the next section, we study the robustness properties of the two Gini covariance matrices along with the Kotz functional through the influence function approach.


\section{Influence function}
The influence function (IF) introduced by Hampel (1974) is a standard
heuristic tool for measuring the effect of infinitesimal perturbations on a
functional $T$. For a distribution $F$ on $\mathbb{R}^d$ and a covariance functional $T: F \mapsto T(F) \in \mathcal{M}^+$ with $\mathcal{M}^+$ being the set of $d \times d$ positive definite matrices, the IF of $T$ at $F$ may be expressed as
\[
\mbox{IF}(\bi{x};T,F)
=\displaystyle \lim_{\varepsilon \rightarrow 0}
\frac{T((1-\varepsilon)F+\varepsilon\delta_{\bi{x}})-T(F)}{\varepsilon},
\;\;\;\bi{x}\in\mathbb{R}^d,
\]
where $\delta_{\bi{x}}$ denotes the point mass distribution at $\bi{x}$. Not only is the IF a local robustness measure of $T(F)$, it is also useful in deriving asymptotic efficiency of the corresponding estimator $T(F_n)$, where $F_n$ is the empirical distribution.
\begin{proposition} \label{thm:ifgc}
The influence function of the Gini covariance matrix $\bi \Sigma_g$ is
$$
IF(\bi x;\bi  \Sigma_g,F)=2\E \frac{(\bi X_1-\bi x)(\bi X_1-\bi x)^T}{\|\bi X_1-\bi x\|}-2\bi \Sigma_g.
$$
\end{proposition}

\begin{remark}
For $d=1$, we obtain the influence function for the Gini mean difference, that is, $IF(x;\sigma_g,F)= 2\E |X_1-x|-2\sigma_g$, which is approximately linear for large $|x|$ in contrast to the quadratic form in $IF(x; \sigma_v^2,F)=\E(X_1-x)^2-\sigma_v^2=(x-\mu)^2$, the influence function of the regular variance.
\end{remark}

The influence function of the affine equivariant GCM is more complicated than that of GCM. Hampel {\em et al.} (1986) showed that, for an affine equivariant scatter functional $M(\cdot)$, the influence function of  $M$ at a spherical distribution $F_0(g)$ in $\mathbb{R}^d$ is given by
\begin{equation}\label{eqn:if}
 IF(\bi x; M, F_0) = \alpha_{M}(\|\bi x\|)\frac{\bi x \bi x^T}{\|\bi x\|^2}-\beta_{ M}(\|\bi x\|)\bi I_{d},
 \end{equation}
where $\alpha_{M}$ and $\beta_{M}$ are two real valued functions depending on $F_0(g)$. Then the influence function of $ M$ at an elliptical distribution $F(\bi \mu, \bi \Sigma, g)$ is
$$IF(\bi x; M, F)=\bi \Sigma^{1/2}IF(\bi \Sigma^{-1/2}(\bi x-\bi \mu); M, F_0)\bi \Sigma^{1/2}.$$

The following corollary states the influence function of the TR version of GCM, which is obtained as a special case of Theorem 2 in Sirki\"{a} {\em et al.} (2007) with $w_1(t)=t$ and $w_2(t)=c(F_0)/d$.

\begin{corollary}\label{thm:ifafgc}
The influence function of the affine equivariant version of the Gini covariance matrix $\bi \Sigma_G$ at a spherical distribution $F_0$ is of the form \eqref{eqn:if} with
\begin{eqnarray*}
\alpha_{\Sigma_G}(\|\bi x\|)&=&\frac{2d(d+2)}{(d+1)c(F_0)}\E\left[ (\left\|\bi X_1-\|\bi x\|\bi e_1\right\|)-\frac{d (\bi X_1)_2^2}{\|\bi X_1-\|\bi x\|\bi e_1\|}\right],\\
\beta_{\Sigma_G}(\|\bi x\|)&=&4-\frac{2d}{(d+1)c(F_0)}\E\left[(\|\bi X_1-\|\bi x\|\bi e_1\|)+\frac{(d+2) (\bi X_1)_2^2}{\|\bi X_1-\|\bi x\|\bi e_1\|}\right],
\end{eqnarray*}
where $(\bi X_1)_2$ denotes the second coordinate of $\bi X_1$, $\bi e_1=(1,0,...,0)^T$, and $c(F_0)=\E_{F_0} \|\bi X_1-\bi X_2\|$. 
\end{corollary}

\begin{remark}
For $d=1$, the influence function for $\sigma_G$ is $IF(x; \sigma_G, F_0)=\alpha_{\sigma_G}(|x|)-\beta_{\sigma_G}(|x|)=4\E|X_1-|x||/c(F_0)-4$, where $c(F_0)=\E_{F_0}|X_1-X_2|$.  Again, it is approximately linear in large values of $|x|$.
\end{remark}

\begin{figure}[tbh]
\centering
\begin{tabular}{ll}
\includegraphics[width=0.49\linewidth]{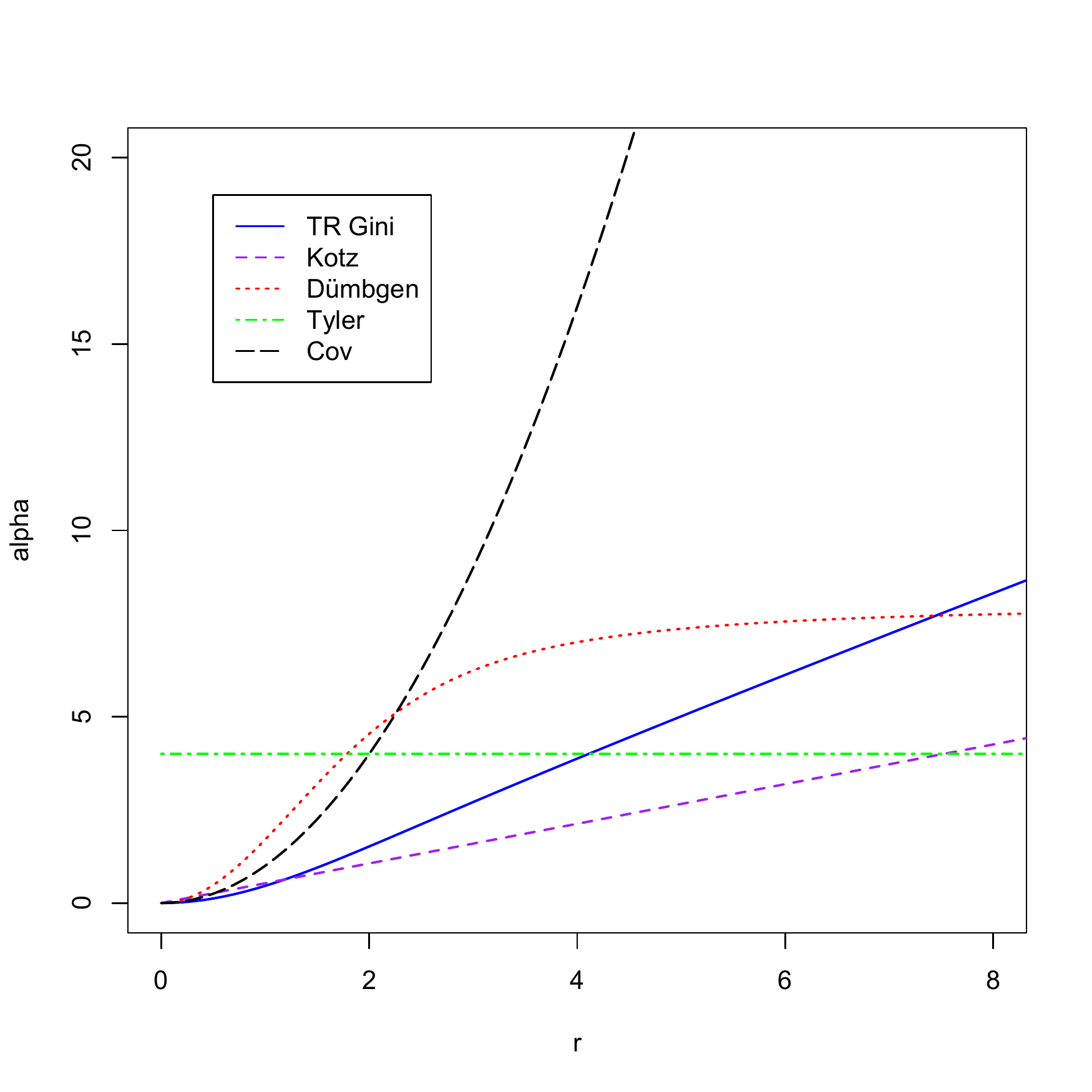}
&
\includegraphics[width=0.49\linewidth]{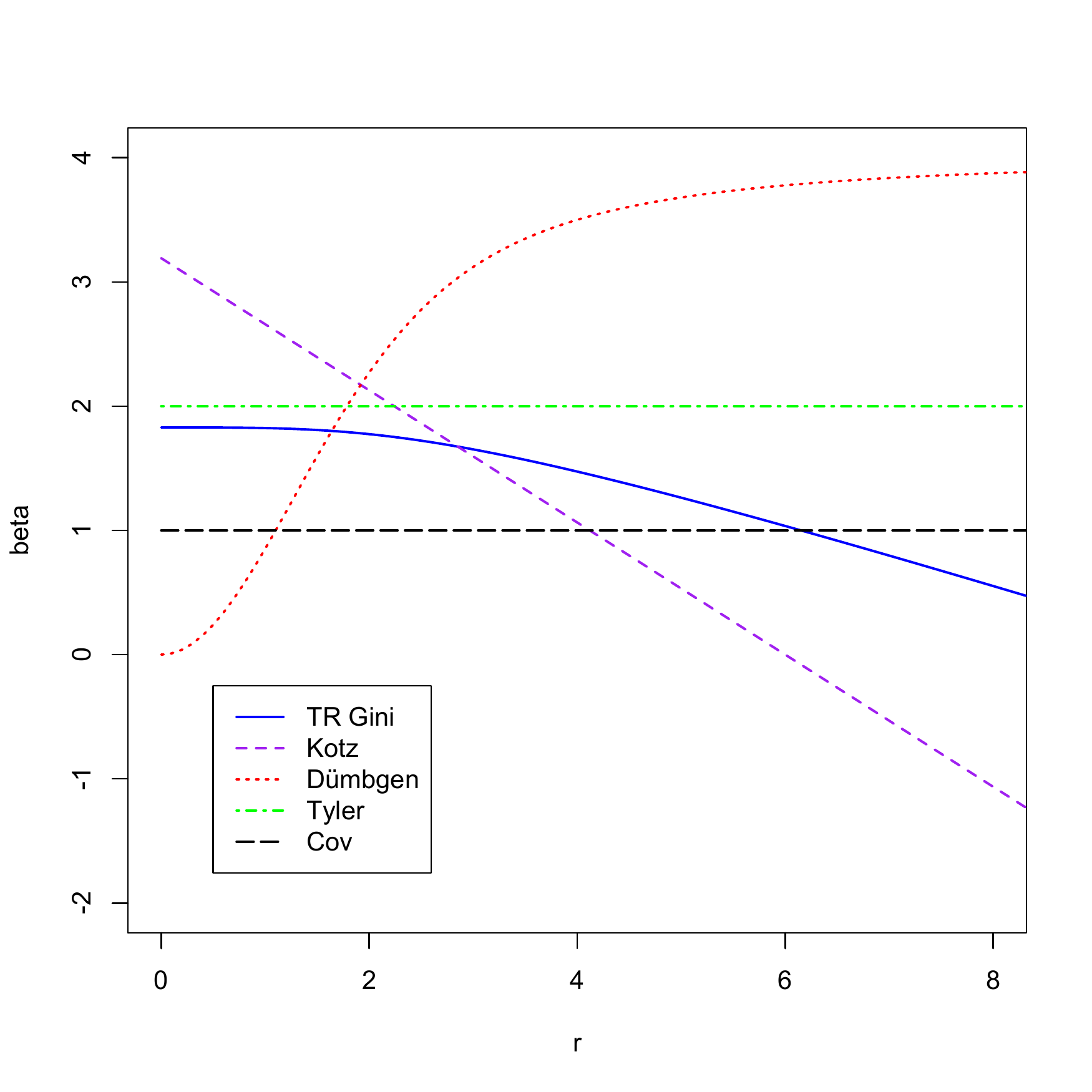}
\end{tabular}
\caption {Functions $\alpha_{M}(r)$ (left panel) and $\beta_{M}(r)$ (right panel) for covariance matrix, Tyler M functional, D\"{u}mbgen functional, Kotz functional and TR Gini covariance matrix under the bivariate standard normal distribution. \label{fig:IF}}
\end{figure}

Applying the result of M-functional from Huber and Ronchetti (2009) (pp. 220-222) to the Kotz functional, we have the following corollary.

\begin{corollary}\label{thm:ifkotz}
The influence function of Kotz functional $\bi \Sigma_K$ at a spherical distribution $F_0$ is of the form \eqref{eqn:if} with
\[
\begin{array}{lcl}
\displaystyle\alpha_{\Sigma_K}(\|\bi x\|)= \frac{d(d+2)}{(d+1)c_1(F_0)}\|\bi x\|,&&
\displaystyle\beta_{\Sigma_K}(\|\bi x\|)=\frac{d}{c_1(F_0)}\left[2-\frac{\|\bi x \|}{d+1}\right],\\
\end{array}
\]
where $c_1(F_0) =\E_{F_0}\|\bi X_1\|$ with $\bi X_1$ from $F_0$. 
\end{corollary}

\begin{remark}
For a spherical distribution $F_0(g)$, $c_1(F_0) = \E r$ where $r$ has the distribution of \eqref{eqn:pdfr}.
\end{remark}

\begin{remark}\label{rem:t}
If $F_0$ is a spherical $t$ distribution ${\cal T}_d(\nu)$ with $\nu>1$, $$c_1(F_0)= \frac{\nu^{1/2}\Gamma[(\nu-1)/2]}{\sqrt{2}\Gamma(\nu/2)}\frac{\sqrt{2}\Gamma[(d+1)/2]}{\Gamma(d/2)}.$$
Note that when $\nu\rightarrow \infty$, using Stirling formula $\Gamma(\nu)\approx \sqrt{2\pi}e^{-\nu}\nu^{\nu-1/2}$, we have $\frac{\nu^{1/2}\Gamma[(\nu-1)/2]}{\sqrt{2}\Gamma(\nu/2)}\rightarrow 1$, which corresponds to the normal case in that $c_1(F_0)=c(F_0)/\sqrt{2}$ as in Remark \ref{rem:normal}.
\end{remark}

\begin{remark}\label{rem:kotz}
If $F_0$ is the spherical Kotz distribution \eqref{eqn:kotz}, then $c_1(F_0)=d$, the mean of Gamma$(d,1)$. 
\end{remark}

Figure \ref{fig:IF} displays functions $\alpha_{M}(r)$ and $\beta_{M}(r)$ for covariance matrix, Tyler M functional, D\"{u}mbgen functional, Kotz functional and TR Gini covariance matrix under the bivariate standard normal distribution.  From \eqref{eqn:if}, the function $\alpha$ is the influence of $\bi x$ on an off-diagonal element of ${ M}$, that is, $IF(\bi x; {M}_{ij},F_0)=\alpha_{M}(\|\bi x\|)u_iu_j$, where $u_i$ and $u_j$ are the $i^{th}$ and $j^{th}$ component of $\bi u=\bi x/\|\bi x\|$.  The influences of diagonal elements of $M$ appear in both $\alpha$ and $\beta$ functions. In other words, $IF(\bi x; {M}_{ii}, F_0) =\alpha_{M}(\|\bi x\|)u_i^2-\beta_{M}(\|\bi x\|)$. This means that for boundedness of the influence at off-diagonal elements, a necessary and sufficient condition is that the $\alpha$ is bounded, while for diagonal elements, one needs boundedness on both $\alpha$ and $\beta$.  As we can see from Figure \ref{fig:IF}, the $\alpha$ and $\beta$ functions of Tyler's and D\"{u}mbgen's M-functionals are bounded. The $\alpha$ function of the covariance matrix is quadratic in the radius $r=\|\bi x\|$, though its $\beta$ function is constant to be bounded. Both functions of the TR Gini covariance matrix are approximately linear for large $r$ and those of the Kotz functionals are linear. This suggests that the TR Gini covariance matrix and Kotz matrix give more protection to moderate outliers than the covariance matrix but they are not robust in the strict sense. The Kotz functional and its symmetrized version TR Gini covariance matrix are $L_1$ methods. They are more robust than $L_2$ methods, and also very efficient (as we will see in the next section). Such properties are also shared by the zonoid scatter matrix (Koshevoy {\em et al.}, 2003), Oja sign and rank covariance matrix (Ollila {\em et al.}, 2003; Ollila {\em et al.}, 2004). They all have influence functions linear or approximately linear in $r$.  

Note that the influence function of the affine equivariant version of spatial rank covariance matrix (MRCM) considered by Visuri {\em et al.} (2000) can not be written as the form of (\ref{eqn:if}) because of the construction way of MRCM with nonlinear transformations.  See Yu {\em et al.} (2015) for more details.


\section{Estimation}
\subsection{Sample Gini Covariance Matrix}

Suppose that ${\cal X}=\{\bi X_1,..., \bi X_n\}$ is a random sample from a continuous distribution $F$ in $\mathbb{R}^d$ and its empirical distribution is $F_n$. Then the sample counterpart of the Gini covariance matrix is obtained by replacing $F$ with the empirical distribution $F_n$ in \eqref{eqn:gcm}. That is,
\begin{align}
\hat{\bi\Sigma}_g=\bi\Sigma_g(F_n)=\frac{2}{n}\sum_{i=1}^n \bi X_i\bi r(\bi X_i)^T 
={n \choose 2}^{-1}\sum_{ i<j}\frac{(\bi X_i-\bi X_j)(\bi X_i-\bi X_j)^T}{\|\bi X_i-\bi X_j\|}.\label{eqn:sgcm}
\end{align}
Clearly, the sample Gini covariance matrix $\bi \Sigma_g(F_n)$ is a matrix-valued $U$-statistic $U_n$ to estimate  $\bi \Sigma_g(F)$ with the kernel $h(\bi x_1,\bi x_2) =(\bi x_1-\bi x_2)(\bi x_1-\bi x_2)^T/\|\bi x_1-\bi x_2\|$.  A straightforward generalization of univariate results on non-degenerated $U$-statistics given in Serfling (1980) establishes $\sqrt{n}$-consistency of $\bi \Sigma_g(F_n)$. This means that for $F$ having a finite second moment,
\begin{align}\label{eqn:rngcm}
\sqrt{n}(\hat{\bi \Sigma}_g-\bi \Sigma_g)=\sqrt{n} (U_n-\bi \Sigma_g)=\sqrt{n}\left[\frac{1}{n}\sum_{i=1}^n IF(\bi X_i; \bi \Sigma_g,F)\right]+\bi R_n,
\end{align}
where the remainder term satisfies $\bi R_n \overset{p} {\to} \bi 0$.  We have the following proposition.

\begin{proposition}\label{thm:gcmnormal}
Let $\bi X_1,...,\bi X_n$ be a random sample from $d$-variate distribution $F$ with a finite second moment. Then $\bi \Sigma_g(F_n)$ is an unbiased, $\sqrt{n}$-consistent estimator of $\bi \Sigma_g(F)$. Furthermore,
$$
\sqrt{n}\, vec (\hat{\bi \Sigma}_g-\bi \Sigma_g) \rightarrow {\cal N}_{d^2}(\bi 0, 4\E[\bi \phi_g(\bi X)\bi \phi_g(\bi X)^T]),
$$
where $\bi \phi_g(\bi x) = vec (\E h(\bi x, \bi X_1) - \bi \Sigma_g)$ with $h(\bi x,\bi x_1) =(\bi x-\bi x_1)(\bi x-\bi x_1)^T/\|\bi x-\bi x_1\|$ and $vec(A)$ stacks columns of $A$ to form a long column vector.
\end{proposition}
Note that $2\bi \phi_g(\bi x)=vec( IF(\bi x;\bi \Sigma_g,F))$. The assumption of a finite second moment guarantees existence of the covariance of the limiting distribution.


\subsection{Sample TR Gini Covariance Matrix}
Replacing $F$ with $F_n$ in \eqref{eqn:trgcm1}, the sample affine equivariant Gini covariance matrix $\hat{\bi \Sigma}_G$ is defined and it is the solution of

\begin{equation}\label{eqn:safgcm}
 \frac{2}{n(n-1)}\sum_{1\leq i <j \leq n}\frac{\hat{\bi \Sigma}_G^{-1/2}(\bi X_i-\bi X_j)(\bi X_i-\bi X_j)^T\hat{\bi \Sigma}_G^{-1/2}}{\sqrt{(\bi X_i-\bi X_j)^T\hat{\bi\Sigma}_G^{-1}(\bi X_i-\bi X_j)}}-\frac{c(F)}{d}\bi I_d =\bi 0.
\end{equation}

The existence and uniqueness of the solution of \eqref{eqn:safgcm} can be established by checking the conditions of scatter M-estimators (Maronna,1976; Huber and Ronchetti, 2009). Those conditions for existence (E) and uniqueness (U) are also used for symmetrized M-estimators in Sirki\"{a} {\em et al.} (2007) and listed below
\begin{description}
\item [E1] $w_1(r)/r^2$ is decreasing, and positive when $r>0$.
\item[E2] $w_2(r)$ is increasing, and positive when $r\geq 0$.
\item [E3] $ w_1(r)$ and $w_2(r)$ are bounded and continuous.
\item[E4] $w_1(0)/w_2(0) < d$.
\item[E5] For any hyperplane $H$, let $P(H)$ be the fraction of pairwise difference belonging to that hyperplane. $P(H)<1-d w_2(\infty)/w_1(\infty)$ and $P(H)\leq 1/d$.
\item [U1] $w_1(r)/r^2$ decreasing.
\item [U2] $w_1(r)$ is continuous and increasing, and positive when $r>0$.
\item [U3] $w_2(r)$ is continuous and decreasing, non-negative, and positive when $0\leq r
<r_0$ for some $r_0$.
\item[U4] For all hyperplane $H$, $P(H)<1/2$.
\end{description}

Our affine equivariant version of Gini covariance estimator is the case with $w_1(t)=t$ and $w_2(t)=c(F)/d$.   It satisfies all except Assumption E3 in which $w_1$ is bounded. However, if we replace E3 with E3',
\begin{description}
\item [E3']  The distribution of $F$ has a finite first moment,
\end{description}
then Lemma 8.3 in Huber and Ronchetti (2009) is still satisfied, hence our estimator does exist and exists uniquely. The assumption E3' is in agreement with the condition of D\"{u}mbgen {\em et al.} (2015) for the Case 1 in which $\psi(t)=\sqrt{t}$.  

Intuitively, we can find the solution of the equation of \eqref{eqn:safgcm}  by a common iterative algorithm:
\begin{equation}\label{eqn:alg}
\hat{\bi\Sigma}_G^{(t+1)} \longleftarrow  \frac{2}{n(n-1)}\frac{d}{c(F)}\sum_{1\leq i <j \leq n}\frac{(\bi x_i-\bi x_j)(\bi x_i-\bi x_j)^T}{\sqrt{(\bi x_i-\bi x_j)^T (\hat{\bi\Sigma}_G^{(t)})^{-1}(\bi x_i-\bi x_j)}}.
\end{equation}
The initial value can take $\hat{\bi\Sigma}_G^{(0)}=\bi I_d$. The iteration stops when $\|\hat{\bi\Sigma}_G^{(t+1)}-\hat{\bi\Sigma}_G^{(t)}\|<\varepsilon$ for a pre-specified number $\varepsilon>0$, where $\|\cdot\|$ can take any matrix norm.  Note that we need to know the distribution $F$ since $c(F)$ is included in (\ref{eqn:alg}). In this case the estimator is Fisher consistent to $\bi\Sigma$. Usually one makes the estimator Fisher consistent at the normal model. That is, one takes $c(F) = 2\Gamma[(d+1)/2]/\Gamma(d/2)$ as stated in Remark \ref{rem:normal}. If one is interested in estimation of correlation matrix or shape matrix (shape matrix is defined later at Section \ref{sec:AE}), there is no need to specify the distribution.  One can delete the factor $d/c(F)$ in the equation of \eqref{eqn:safgcm} and obtain its solution for estimation of scatter matrix up to a factor. 

The above algorithm is called the fixed-point algorithm and its convergence from any start points has been rigorously proved (Tyler, 1987). However, it can be rather slow for high dimensions and large sample sizes. The very recent paper by D\"{u}mbgen, Nordhausen and Schuhmacher (2016) provide much faster new algorithms by utilizing a Taylor expansion of second order of the target functional. For large $n$, they approximate symmetrized estimators by considering the surrogate ones rather than all pairwise differences. Based on their idea, an algorithm for our Gini estimator can be developed and added to their R package ``fastM" (D\"{u}mbgen {\em et al.}, 2014). 

If we assume that the location parameter $\bi \mu$ is known, then the MLE of $\bi \Sigma$ in the Kotz distribution \eqref{eqn:kotz} is found to be a scatter M-estimator, which is the solution of $\hat{\bi \Sigma}$ in the equation below:
\begin{align}
\frac{1}{n}\sum_{i=1}^n\frac{(\bi X_i-{\bi \mu})(\bi X_i-{\bi \mu})^T} {\sqrt{(\bi X_i-{\bi\mu})^T\hat {\bi \Sigma}^{-1}(\bi X_i-{\bi \mu})}}= \hat{\bi \Sigma}.\label{eqn:Kotz_Sigma}
\end{align}
The solution of $\hat{\bi \Sigma}$ in \eqref{eqn:Kotz_Sigma} is denoted as $\hat {\bi \Sigma}_K$, which is ${\bi \Sigma}_K(F_n)$.  Assuming a known location parameter is for avoiding some restrictive regularity conditions for the simultaneous M-estimators.  The simultaneous one is treated in Roelant and Van Aelst (2007) and Arslan (2010). Our TR version GCM estimator is the symmetrized scatter MLE of the Kotz distribution without the need of reference to the location parameter, and hence avoids the above situation. 

D\"{u}mbgen {\em et al.} (2015) provided a general treatment and asymptotics for M-estimation of multivariate scatter. The Kotz and TR Gini estimators are examples of their Case 1. Hence by using their Theorem 6.11, $\sqrt{n}$-consistency of $\hat {\bi \Sigma}_G$ and $\hat {\bi \Sigma}_K$ under a spherical distribution $F_0$ is established as follows. 

\begin{proposition} \label{thm:consistent_trgcm}
Let $\bi X_1,...,\bi X_n$ be a random sample from a spherical distribution $F_0$ in $\mathbb R^d$.  Under the assumption of finite second moment of $F_0$, $\hat{\bi \Sigma}_G$ is $\sqrt{n}$-consistent estimator of $\bi \Sigma_G(F_0)=\bi I_d$ and $\hat{\bi \Sigma}_K$ is $\sqrt{n}$-consistent estimator of $\bi \Sigma_K(F_0)=[c_1(F_0)]^2/d^2 \bi I_d$, where $c_1(F_0)=\E_{F_0}[\|\bi X\|]$.
\end{proposition}

\begin{remark}
If $F_0$ is the spherically distributed Kotz distribution, $\bi \Sigma_K(F_0)=\bi I_d$ and both $\hat{\bi \Sigma}_G$ and $\hat{\bi \Sigma}_K$ are consistent scatter estimators.
\end{remark}

Once we obtain the $\sqrt{n}$-consistency of $\hat{\bi \Sigma}_G$, we are able to use Theorem 4 of Sirki\"{a} {\em et al.} (2007), in which they assume $\sqrt{n}$-consistency of symmetrized M-estimators to establish asymptotic normality. In the following we give the result for our estimator.

\begin{corollary}\label{thm:normafgcm}
Let $\bi X_1,...,\bi X_n$ be a random sample from a spherical distribution $F_0$ in $\mathbb R^d$. If the covariance matrix (second moments) of $F_0$ exists, then
$$
\sqrt{n}\; vec(\hat{\bi \Sigma}_G-\bi I_{d})\rightarrow N_{d^2}(\bi 0, \E[ vec(IF(\bi X;\bi \Sigma_G,F_0))vec(IF(\bi X;\bi \Sigma_G,F_0))^T]).
$$
\end{corollary}
According to \eqref{eqn:if} and Corollary \ref{thm:ifafgc}, the covariance matrix of the limit distribution
$\E[ vec(IF(\bi X;\bi \Sigma_G,F_0))vec(IF(\bi X;\bi \Sigma_G,F_0))^T]$ can be written as
$$
ASV(\hat{\bi \Sigma}_{G_{12}};F_0)(\bi I_{d^2}+\bi 1_{d,d})+ASC(\hat{\bi \Sigma}_{G_{11}},\hat{\bi \Sigma}_{G_{22}};F_0)vec(\bi I_{d})vec(\bi I_{d})^T,
$$
where $\bi 1_{d,d}$ is $d^2\times d^2$ matrix with $(i,j)$-block being equal to a $d\times d$ matrix that has 1 at entry $(j,i)$ and 0 elsewhere. $ASV(\hat{\bi \Sigma}_{G_{12}};F_0)$ denotes the asymptotic variance of an off-diagonal element and $ASC(\hat{\bi \Sigma}_{G_{11}},\hat{\bi \Sigma}_{G_{22}};F_0)$ denotes the covariance of any two diagonal elements. With Corollaries \ref{thm:normafgcm} and \ref{thm:ifafgc}, we have

\begin{align}
&ASV(\hat{\bi \Sigma}_{G_{12}};F_0)
\nonumber\\
&= \frac{4d(d+2)}{(d+1)^2c^2(F_0)}\E\left[\E(\|\bi X_1-\|\bi X_2\|\bi e_1\| -\frac{d(\bi X_1)_2^2}{\|\bi X_1-\|\bi X_2\|\bi e_1\|})|\bi X_2\right]^2; \label{eqn:asvG}\\
&ASV(\hat{\bi \Sigma}_{G_{11}};F_0) \nonumber\\
&=\frac{2(d-1)}{d}ASV(\hat{\bi \Sigma}_{G_{12}};F_0)+16\left[\frac{\E\left[\E(\|\bi X_1-\|\bi X_2\|\bi e_1\|)|\bi X_2\right]^2}{c^2(F_0)}-1\right]; \nonumber\\
&ASC(\hat{\bi \Sigma}_{G_{11}},\hat{\bi \Sigma}_{G_{22}};F_0)=ASV(\hat{\bi \Sigma}_{G_{11}};F_0)-2ASV(\hat{\bi \Sigma}_{G_{12}};F_0).\nonumber
\end{align}

Using the affine equivariance property of $\hat{\bi \Sigma}_G$ and Kronecker product $\otimes$, the limiting distribution of $\sqrt{n}\;vec(\hat{\bi \Sigma}_G-\bi \Sigma)$ at the elliptical distribution $F$ is multivariate normal with zero mean and covariance matrix
\begin{equation}\label{eqn:ASVF}
ASV(\hat{\bi \Sigma}_{G_{12}};F_0)(\bi I_{d^2}+\bi 1_{d,d})(\bi \Sigma \otimes \bi \Sigma)+
ASC(\hat{\bi \Sigma}_{G_{11}},\hat{\bi \Sigma}_{G_{22}};F_0)vec(\bi \Sigma)vec(\bi \Sigma)^T.
\end{equation}

Checking the conditions (N1-N4) of MLE proposed by Huber (1967), we are able to establish the normality of Kotz estimator $\hat{\bi \Sigma}_K$ assuming a known location parameter.
\begin{proposition}\label{thm:normkotz}
Let $\bi X_1,...,\bi X_n$ be a random sample from spherical distribution $F_0(g)$ in $\mathbb R^d$. If the second moment of $F_0(g)$ exists and the first moment is known, then
$$
\sqrt{n}\; vec(\hat{\bi \Sigma}_K-\bi \Sigma_K)\rightarrow N_{d^2}(\bi 0, \E[ vec(IF(\bi X;\bi \Sigma_K,F_0))vec(IF(\bi X;\bi \Sigma_K,F_0))^T]).
$$
\end{proposition}
With the results of  Corollary \ref{thm:ifkotz} and Proposition \ref{thm:normkotz}, we have
\begin{equation}\label{eqn:asvkotz}
\displaystyle ASV(\hat{\bi \Sigma}_{K_{12}};F_0)= \frac{d(d+2) \E_{F_0}[ \|\bi X\|^2]}{(d+1)^2 [c_1(F_0)]^2}, 
\end{equation}
in which $\E_{F_0}[ \|\bi X\|^2]=\E R^2$ with $R$ having the distribution of (\ref{eqn:pdfr}). 

\subsection{Asymptotic Efficiency}\label{sec:AE}
Although our TR Gini covariance estimator is Fisher consistent to the scatter matrix since it is corrected by $c(F_0)/d$,  we consider its shape estimator in order to compare its limiting efficiency with that of the Tyler and D\"{u}mbgen M-estimators. The shape matrix associated with the scatter functional $\bi \Sigma$ is
$$ W(F)=\frac{d}{Tr(\bi \Sigma(F))}\bi \Sigma(F).$$
Note that there are also other definitions for a shape matrix. For example, Paindaveine (2008) uses the determinant. Here we use the shape matrix based on the matrix trace because it allows us to compare asymptotic efficiency more easily. Tyler and D\"{u}mbgen estimators estimate the shape matrix. At elliptical distributions, all shape estimators estimate the same population quantity and hence are comparable without any correction factors.  Theorem 5 of Sirki\"{a} {\em et al.} (2007) states that
a single number characterizes the limiting distribution of the shape estimators at $F_0$ and that number  is the variance of off-diagonal elements of $\hat{\bi \Sigma}$ or $\hat{W}$,  $\tau$. In general, the asymptotic relative efficiency (ARE) of an estimator $T_1$ with respect to another estimator $T_2$ is defined as the ratio of $ASV(T_2)$ and $ASV(T_1)$.   Hence for shape estimators, the ARE of $\hat{W}_1$ with respect to $\hat{W}_2$ is $\tau( \hat{W}_2)/\tau( \hat{W}_1)$. 

\begin{table}[thb]
\centering
\begin{tabular}{llcccccr}\toprule\toprule
&&${\cal T}_d(5)$&${\cal T}_d(6)$&${\cal T}_d(8)$&${\cal T}_d(15)$&${\cal T}_d(\infty)$&$Kotz(d)$\\  \midrule
\multirow{4}{*}{$d=2$} & Tyler& 1.50&1.00&0.75&0.59&0.50&0.83\\
& D\"umbgen & 2.36&1.57&1.26&1.01&0.91&1.22 \\
& Kotz&2.25 &1.56&1.22&1.00&0.88&1.25\\
& TR Gini&2.09&1.48&1.24&1.05&0.98&1.21\\ 
& Zonoid&2.00 &1.45& 1.18 &1.03 &0.96&1.11\\ \midrule
\multirow{4}{*}{$d=3$}& Tyler& 1.80&1.20&0.90& 0.71& 0.60& 0.90\\
& D\"umbgen  & 2.38& 1.66& 1.27& 1.04& 0.92&1.18 \\
& Kotz&2.31 &1.60&1.25&1.03&0.91&1.20\\
&TR Gini&2.14&1.53&1.25&1.06&0.99&1.17\\  
& Zonoid& 1.96 &1.43 &1.18& 1.04&0.97&1.07\\ \midrule
\multirow{4}{*}{$d=4$}&Tyler& 2.00&1.33&1.00&0.79&0.67&0.93\\
&D\"umbgen & 2.39& 1.69&1.30& 1.06&0.93 &1.15\\
& Kotz&2.34 &1.63&1.27&1.05&0.92&1.17\\
& TR Gini&2.21&1.56&1.26&1.09&0.99&1.15\\  
& Zonoid& 1.93 &1.41& 1.17 &1.04&0.98&1.05 \\ \midrule
\multirow{4}{*}{$d=5$}&Tyler& 2.14&1.43&1.07&0.84&0.71&0.95\\
&D\"umbgen & 2.50& 1.71&1.31& 1.07& 0.94&1.13\\
& Kotz&2.37 &1.65&1.29&1.06&0.93&1.14\\
&TR Gini&2.28&1.57&1.26&1.09&0.99&1.11\\ 
& Zonoid& 1.91& 1.40 &1.17& 1.04&0.99&1.04 \\ \bottomrule\bottomrule
\end{tabular}
\caption{Asymptotic relative efficiencies of the shape estimators based on the Tyler M-estimator, D\"umbgen, Kotz M-estimator, TR Gini covariance estimator and Zonoid covariance estimator relative to the regular shape estimator at different distributions $F_0$ at different  $d$-dimension.}
\label{tbl:ARE}
\end{table}

Listed in Table \ref{tbl:ARE} are the limiting efficiencies of shape estimators with respect to the shape estimator based on the regular sample covariance matrix (i.e. the regular shape estimator). The efficiencies are considered under spherical Kotz($d$) distribution and ${\cal T}_d(\nu)$ distributions at different dimensions $d$ with different degrees of freedom $\nu$, with $\nu=\infty$ referring to the normal case.  The variance of the off-diagonal element of the regular shape estimator at $F_0$ equal to $1+\kappa(F_0)$, where $\kappa(F_0)$ is the kurtosis of $F_0$. That is, $\tau$ of the regular shape estimator is $(\nu-2)/(\nu-4)$ in the ${\cal T}_d(\nu)$-distributions for $\nu>4$ and $(d+3)/(d+1)$ in the $Kotz(d)$ distribution (Wang, 2009; Zografos, 2008).  In the normal case, $\tau=1$ corresponds to that of the ${\cal T}_d(\nu)$-distribution case when $\nu\rightarrow \infty$.  $\tau$ of the Tyler estimator is always $(d+2)/d$ for any distribution in $\mathbb{R}^d$. From (\ref{eqn:asvkotz}),  the asymptotic variance of off-diagonal elements of  the Kotz shape estimator under $F_0$ is equal to $d(d+2)\E[\|\bi X\|^2]/((d+1)^2[\E\|\bi X\|]^2)$ with $\bi X$ from $F_0$. For example, ASV of the Kotz shape estimator under the $Kotz(d)$ distribution is $(d+2)/(d+1)$. The variances of off-diagonal elements of the TR Gini shape estimator are given by (\ref{eqn:asvG}), and computed through a combination of numerical integration and Monte Carlo simulation. More specifically, for $d=2$, the inner expectation of (\ref{eqn:asvG}) is computed by a double integration and the outer expectation is estimated by an empirical mean on a sample of size $10^8$. For $d >2$, all calculations are through simulations on samples with size $10^8$. The asymptotic variance of off-diagonal elements of the zonoid shape estimator under $F_0$ is $d(4\E[\|\bi X\|^2]-3[\E\|\bi X\|]^2)/((d+2)[\E\|\bi X\|]^2)$ (Koshevoy {\em et al.}, 2003). For example, under the Kotz distributions, the ASV of the zonoid shape estimator is $(d+4)/(d+2)$.

From Table \ref{tbl:ARE}, it can be seen that the ARE of each shape estimator decreases as $\nu$ increases in ${\cal T}_d(\nu)$ distributions, and the ARE of Tyler, D\"umbgen, Kotz and TR Gini shape estimators increases as dimension $d$ increases.  In the normal cases, TR Gini estimator has a 98\% ARE for $d=2$ and 99\% for $d\geq 3$. With very little loss in efficiency in the normal case, the TR Gini estimator gains efficiency in the heavy tailed distributions. For example, its ARE is greater than 2 relative to the regular shape estimator in the ${\cal T}_d(5)$ distribution.  The Tyler estimator has the lowest ARE among all estimators  except the Zonoid estimator for all distributions considered. In particular, the symmetrized D\"umbgen estimator is more efficient than its counterpart, the Tyler estimator, in all distributions.  However, such a result does not hold for all symmetrized estimators. TR Gini shape estimator is more efficient than Kotz estimator in ${\cal T}_d(15)$ and ${\cal T}_d(\infty)$, but less efficient in the Kotz and ${\cal T}_d(\nu)$ distributions with $\nu=5,6$. It is worthwhile to point out that Gerstenberger and Vogel (2015) studied efficiency of Gini mean difference. Their results complement ours for Kotz and TR Gini estimator when $d=1$.  The ARE's
of the zonoid shape estimator under ${\cal T}_d(\infty)$ are 0.96, 0.97, 0.98 and 0.99, respectively for $d=2,3,4,5$. Under ${\cal T}_2(\nu)$, their ARE's are 2.00, 1.45, 1.18 and 1.03, respectively for $\nu=5,6,8,15$. Those numbers are similar to (slightly smaller than) the ARE's of our TR Gini shape estimator, which is not surprising since both are multivariate extensions of the mean deviation or mean difference and both have linear or approximately linear influence functions. They are highly efficient at the normal and fairly robust at the heavy-tailed cases. For ${\cal T}_d(5)$,  ${\cal T}_d(6)$, ${\cal T}_d(8)$  and Kotz distributions, the efficiency of the Zonoid shape estimator decreases with $d$, which is different from other estimators. At ${\cal T}_5(5)$, the Zonoid shape estimator is least efficient among M-estimators and symmetrized M-estimators, but it is much efficient than the regular shape estimator. 

\subsection{Finite Sample Efficiency}

We conduct a small simulation to study finite sample efficiencies of the shape estimators with respect to the regular shape estimator. $M=10000$ samples of two different sample sizes ($n=50,200$) at two different dimensions ($d=2,5$) are drawn from spherical ${\cal T}$-distributions with 5, 8 and $\infty$ degrees of freedoms and from spherical Kotz distribution. We use R Package ``mnormt" (Azzalini and Genz, 2016) to generate samples from multivariate ${\cal T}$-distributions and normal distribution. We generate a random vector $\bi X$ from spherical Kotz distribution by  $\bi X=R \bi U$, in which $R$ is distributed from the Gamma distribution with the shape parameter being $d$ and the scale parameter being 1 and $\bi U=\bi Z/\|\bi Z\|$ with $\bi Z$ being a vector formed by $d$ iid standard normal variables.  If a random sample from Kotz$(\bi \mu,\bi \Sigma)$ is required, then by taking $\bi \Sigma$'s Cholesky decomposition $L$,  we have $\bi Y =L\bi X +\bi \mu $ from Kotz$(\bi \mu,\bi \Sigma)$. 

\begin{table}[ht]
\centering
\small
\begin{tabular}{@{}ll|ccccccccccr@{}}\toprule\toprule
&&\multicolumn{2}{c}{${\cal T}_d(5)$}&&\multicolumn{2}{c}{${\cal T}_d(8)$}&&\multicolumn{2}{c}{${\cal T}_d(\infty)$}&&\multicolumn{2}{r}{$Kotz(d)$}\\
\cmidrule(l){3-4}\cmidrule(l){6-7}\cmidrule(l){9-10}\cmidrule(l){12-13}
&$n \backslash d$ &$2$ & $5$ && $2$ & $5$ && $2$ & $5$&&$2$ & $5$ \\
\midrule
Tyler &50 &0.81&	1.12&&0.61&	0.71&&0.45&0.49	&&  0.71&0.75\\
        & 200 &1.14&1.60&&	0.82&	1.01&&0.59&	0.69& & 0.79&	0.90\\
&$\infty$ & 1.50&	2.14&&	0.75&	1.07&&	0.50&	0.71& & 0.83&	0.95\\ \midrule

D\"umbgen&50&1.27&	1.73&&	1.02&	1.15&&	0.83&	0.89 && 1.04&0.94\\
&200 &1.35&	1.88&&	1.03&	1.23&&	0.81&	0.91&& 1.17&	1.09\\
&$\infty$ & 2.36&	2.50&&	1.26&	1.31&&	0.91&	0.89& & 1.22&	1.13\\ \midrule

Kotz&50&1.41&	1.72&&	1.15&	1.19&&	0.91&	0.96 && 1.23&1.13\\
&200&1.54&	1.87&&	1.22&	1.27&&	0.95&	0.94& & 1.24&	1.14\\
&$\infty$ & 2.25&	2.37&&	1.22&	1.29&&	0.88&	0.93& & 1.25&	1.14\\ \midrule

TR Gini&50&1.31&	1.60&&	     1.14&	1.18&&	0.98&	0.99& & 1.15&1.09\\
& 200&1.36&	1.67&&	1.16&	1.21&&	0.99&	0.99& & 1.18&	1.10\\
&$\infty$ & 2.09&	2.28&&	1.24&	1.26&&	0.98&	0.99& & 1.21&	1.11\\ \midrule

MRCM & 50 &0.95& 1.17&&0.72&0.71&&0.52&0.49&&0.78&0.75 \\
& 200&1.29&1.50&&0.92&0.92&&0.63&0.60&&0.84&0.81 \\
$(Q_n)$ &200&1.65&1.83&&1.11&1.19&&0.83&0.87&&1.07&1.04\\

\bottomrule\bottomrule
\end{tabular}
\caption{
Finite sample relative efficiencies of the shape estimators with respective to the regular shape matrix at different distributions $F_0$. }
\label{tbl:FRE}
\end{table}

In the simulation, all M-estimators and symmetrized M-estimators are calculated by the fixed-point algorithm. Tyler and Kotz shape estimators use the true location values in the computation. {\em tyler.shape} and {\em duembgen.shape} functions in R package ``ICSNP" (Nordhausen{\em et al.}, 2015) are used for computing Tyler and D\"umbgen estimators. Also {\em spatial.rank} function of ``ICSNP" is used for TR Gini shape estimator. The convergence criterion uses Frobenius matrix norm with $\varepsilon$ being the default value $10^{-6}$ and the maximum number of iterations setting to be 100. We also include the affine equivariant spatial rank shape estimator (MRCM) for comparison. It uses the median of absolute deviation (MAD) as univariate scale estimator. An alternative to MAD, $Q_n$, is also included to see efficiency improvements of MRCM. The Zonoid shape estimator is not included in the finite sample efficiency comparison study due to its high computation complexity $O(n^{d+1})$. 

For each estimator, the mean squared errors of off-diagonal elements are computed. That is, $$MSE(\hat{\bi \Sigma}_{ij}) =\frac{1}{M}\sum_{m=1}^{M}(\hat{\bi \Sigma}_{ij}^{(m)}-\bi \Sigma_{ij})^2$$ for $i\neq j$. Obviously, here we have $\bi \Sigma_{ij}=\bi I_{ij}=0$. Since the off-diagonal elements have equal variances and are uncorrelated,  the average of their MSEs is computed. The finite sample relative efficiencies listed in Table \ref{tbl:FRE} are ratios of the mean MSE of the regular shape matrix to that of each estimator. The asymptotic relative efficiencies ($n=\infty$) from Table \ref{tbl:ARE} are also listed in Table \ref{tbl:FRE} for convenient reference.

The results of finite sample study show that Kotz and TR Gini estimators have a relatively fast convergence to their limiting efficiencies. Even for $n=50$ of the normal and Kotz cases, their finite sample efficiencies are already close to the asymptotic ones.  For the Tyler estimator, the convergence is slower, and the loss in efficiency is larger for finite sample sizes comparing to that of others.
In the case of the ${\cal T}(5)$ distribution, the convergence to the limiting efficiency is much slower than that of the other cases. Low efficiency of MRCM can be explained by low efficiency of the univariate scale estimator MAD. Improvement can be done by using other robust alternatives which are more efficient, as suggested by Rousseeuw and Croux (1993).  They recommended $Q_n$ which is given by the 0.25 quantile of the pairwise distances multiplying some correction factor. For the normal distribution under the size $n=200$, if $Q_n$ is used, the RE of MRCM increases to 0.83 for $d=2$ and 0.87 for $d=5$. Similar improvements are observed for other distributions also. 


\section{Conclusion}

We have extended the univariate Gini mean difference to the multivariate case and proposed two versions of Gini covariance matrix (GCM).  New covariance matrices are based on pairwise differences. Thus the location center needs not be estimated nor known. Their properties have been explored. They possess the block independence property, which allow them beneficial in many applications. Their influence functions have been derived. It was found that the influence functions of GCM are approximately linear, which is unbounded. In a strict sense, they are not highly robust. However, they are highly efficient under normal distributions. They have greater than 98\% asymptotic relative efficiency with respect to sample covariance matrix. On the other hand,  they are more robust than the covariance matrix which has influence function of a quadratic form. GCM will give more protection to moderate outliers than the covariance matrix.  Similar properties are also shared by the Oja sign or rank covariance matrix and the zoniod or lift-zoniod covariance matrix, but our proposed ones enjoy computational ease. Hence the proposed affine equivariant GCM provides us an option for estimating scatter matrix with a consideration to balance well among efficiency, robustness and computation. 


\section{Appendix}

\noindent {\bf Proof of Theorem \ref{thm:eigv}}. We first show that $\bi X_1-\bi X_2$ is elliptically distributed with center $\bf 0$ and scatter parameter $2\bf \Sigma$ by its characteristic function as follows. 
$$ \E e^{i\bi t^T(\bi X_1-\bi X_2)}= \E e^{i\bi t^T\bi X_1}\E e^{-i\bi t^T\bi X_2}= e^{i\bi t^T\bi \mu}e^{-i\bi t^T\bi \mu}\psi^2(\bi t^T\bi \Sigma \bi t):=\psi^*(2\bi t^T\bi \Sigma \bi t),$$
where $\psi^*(s) = \psi^2(s/2)$. Note that except for normal distributions, $\bi X_1-\bi X_2$ has a different generating function $g^*$ from $g$, the one for $\bi X$.  

Let $\bi Z_i=V^T(\bi X_i-\bi\mu)$ for $i=1,2$,  then $\bi Z_1-\bi Z_2=V^T(\bi X_1-\bi X_2)$ follows a centered elliptical distribution with diagonal scatter matrix $2\Lambda$. We can write $(2\Lambda)^{-1/2}(\bi Z_1-\bi Z_2)= R\bi U$ with $R=\|(2\Lambda)^{-1/2}(\bi Z_1-\bi Z_2)\|$ and $\bi U=(2\Lambda)^{-1/2}(\bi Z_1-\bi Z_2)/R$ being independent with $R$ and uniformly distribution on the unit sphere. Then
$$
\begin{array}{lcl}
\displaystyle\bi\Sigma_g&=&\displaystyle\E\left[\frac{(\bi X_1-\bi X_2)(\bi X_1-\bi X_2)^T}{\|\bi X_1-\bi X_2\|}\right]=V\E\left[\frac{(\bi Z_1-\bi Z_2)(\bi Z_1-\bi Z_2)^T}{\|\bi Z_1-\bi Z_2\|}\right]V^T\\\\
&=&\displaystyle V\E\left[\frac{2R^2\Lambda^{1/2}\bi U\bi U^T\Lambda^{1/2}}{\sqrt{2R^2\bi U^T\Lambda^{1/2}\Lambda^{1/2}\bi U}}\right]V^T=\sqrt{2}\E RV\E\left[\frac{\Lambda^{1/2}\bi U\bi U^T\Lambda^{1/2}}{\sqrt{\bi U^T\Lambda\bi U}}\right]V^T.
\end{array}
$$
Denote $\sqrt{2}\E R$ as $c(F)$,  the proof is complete.
\hfill$\square$

\vspace{2ex}
\noindent{\bf Proof of Theorem \ref{thm:afeq}}. Multiplying $A$ on the left and $A^T$ on the right to both sides of Equation (\ref{eqn:trgcm2}), we have
$$
A\bi \Sigma_G A^T = \frac{d}{c(F)}\E \frac{A(\bi X_1-\bi X_2)(\bi X_1-\bi X_2)^TA^T}{\sqrt{(\bi X_1-\bi X_2)^T\bi \Sigma_G^{-1}(\bi X_1-\bi X_2)}}.
$$
Since $A$ is nonsingular, $A^{-1}$ and $(A^T)^{-1}$ exist. Hence
$$A\bi \Sigma_G A^T= \frac{d}{c(F)}\E \frac{A(\bi X_1-\bi X_2)(\bi X_1-\bi X_2)^TA^T}{\sqrt{(\bi X_1-\bi X_2)^TA^T(A\bi \Sigma_GA^T)^{-1}A(\bi X_1-\bi X_2)}}.
$$
It means that $A\bi\Sigma_G A^T$ is the TR version of Gini covariance matrix for $A\bi X+\bi b$, where $\bi X$ is random vector from distribution $F$. \hfill $\square$

\vspace{2ex}

\noindent{\bf Proof of Proposition \ref{thm:ifgc}}. The proof is straightforward. Let $\bi Y_1$ and $\bi Y_2$ be independently distributed from $F_{\varepsilon}=(1-\varepsilon)F+\varepsilon \delta_{\bi x}$ and $\bi X_1$ and $\bi X_2$ independently distributed from $F$, then we have
\begin{align*}
&\bi \Sigma_g(F_{\varepsilon}) = \E_{F_\varepsilon}(\bi Y_1-\bi Y_2)\bi s(\bi Y_1-\bi Y_2)^T \\
&= (1-\varepsilon)^2 \E_F\frac{(\bi X_1-\bi X_2)(\bi X_1-\bi X_2)^T}{\|\bi X_1-\bi X_2\|}+
2\varepsilon(1-\varepsilon)\E_F \frac{(\bi X_1-\bi x)(\bi X_1-\bi x)^T}{\|\bi X_1-\bi x\|}.
\end{align*}
Then the result for $IF(\bi x;\bi \Sigma_g, F)$ follows.  \hfill $\square$

\vspace{2ex}
\noindent{\bf Proof of Corollary \ref{thm:ifafgc}}.
The affine equivariant version of Gini covariance matrix $\bi \Sigma_G$ is a symmetrized M-functional with $w_1(t)=t$ and
$w_2(t)=c(F)/d=c(F_{0})/d$. From Theorem 2 of Sirkia {\em et al.} (2007), we get
\begin{eqnarray*}
\eta_1&=&\frac{(d+1)\E[\|\bi X_1- \bi X_2\|]}{2d(d+2)}=\frac{(d+1)c(F_{0})}{2d(d+2)},\\[1ex]
\eta_2&=&\frac{\E\|\bi X_1- \bi X_2\|}{4d}=\frac{c(F_{0})}{4d}.
\end{eqnarray*}
Thus, the result is obtained.
\hfill $\square$

\vspace{2ex}

\noindent{\bf Proof of Corollary \ref{thm:ifkotz}}.
We have the influence function of M-functional $M$ in the form $IF(\bi x; M, F_0)=-2\dot{W}$ where $W=M^{-1/2},$ $\dot{W}=IF(\bi x;W, F_0),$ and 
\begin{eqnarray*}
\frac{1}{d}tr(W)&=&-\frac{\frac{1}{d}w_1(\|\bi x\|)-w_2(\|\bi x\|)}{\E[(\frac{1}{d}w^{'}_1(\|\bi Y\|)-w^{'}_2(\|\bi Y\|))\|\bi Y\|]},\\
\dot{W}-\frac{1}{d}tr(W)\bi I_d&=&-\frac{d+2}{2}\frac{w_1(\|\bi x\|)(\frac{\bi x\bi x^T}{\|\bi x\|^2}-\frac{1}{d}\bi I_d)}{\E_{y}[w_1(\|\bi Y\|)+\frac{1}{d}w^{'}_1(\|\bi Y\|)\|\bi Y\|]},
\end{eqnarray*}
where $\bi Y$ is a random vector from the distribution $F_0$ (see pages 220-222 of Huber and Ronchetti (2009)). 

With $w_1(t)=t$  and $w_2(t)=1$ along with $w_1^{'}(t)=1$ and $w_2^{'}(t)=0$ for $\bi \Sigma_{K},$ solving for $\dot{W}$ in the above equations we get
\begin{eqnarray*}
\dot{W}&=&\frac{-d(d+2)}{2(d+1)\E\|\bi Y\|}\|\bi x\|\frac{\bi x\bi x^T}{\|\bi x\|^2}+\frac{d+2}{2(d+1)\E\|\bi Y\|}\|\bi x\|\bi I_d-\frac{\|\bi x\|-d}{\E\|\bi Y\|}\bi I_d.
\end{eqnarray*}
Let  $c_1(F_0)=\E\|\bi Y\|.$ Therefore, we obtain
\begin{align*}
&IF(\bi x; \bi \Sigma_{K},F_0)=-2\dot{W}\\
&=\frac{d(d+2)}{(d+1)c_1(F_0)}\|\bi x\|\frac{\bi x\bi x^T}{\|\bi x\|^2}+\frac{d}{(d+1)c_1(F_0)}\|\bi x\|\bi I_d-\frac{2d}{c_1(F_0)}\bi I_d.\\
\end{align*}
Hence the result follows.
\hfill $\square$

\vspace{2ex}

\noindent{\bf Proof of Proposition \ref{thm:gcmnormal}}. We only prove the asymptotic normality result. 

The normality of an U-statistic follows from the central limit theorem on its first order Hoeffding decomposition provided that the U-statistic is non-degenerated.  Here we need to show that  $\E[\bi \phi_g(\bi X)\bi \phi_g(\bi X)^T] >\bi 0$ and exists. The existence is guaranteed by the assumption of finite second moment.  Hence it is sufficient to  prove that $\phi_g(\bi X)$ is of full rank almost everywhere. This is true if $P(\bi X \in V)=0$ for any proper linear subspace $V$ ($dim(V)<d$). Particularly, this is true for continuous distribution $F$. \hfill $\square$

\vspace{2ex}

\noindent{\bf Proof of Proposition \ref{thm:consistent_trgcm}}.  Kotz and TR Gini estimators are examples of the Case 1 considered in D\"{u}mbgen {\em et al.} (2015) with the symmetrization order 1 and 2, respectively. Using the same notations of D\"{u}mbgen {\em et al.} (2015),  Kotz and TR Gini estimators are the cases with  $\rho(s) = \sqrt{s}, \psi(s) = \frac{1}{2} \sqrt{s}$ and $\psi_2(s)= \frac{1}{4}\sqrt{s}$, which satisfy all conditions on $\rho, \psi$ and $\psi_2$. Under continuous distribution $F_0$ with finite second moments,  Theorem 6.11 holds for Kotz and TR Gini estimators, and hence they are $\sqrt{n}$ consistent to $\bi \Sigma_K$ and $\bi \Sigma_G$, respectively. \hfill $\square$

\vspace{2ex}

\noindent{\bf Proof of Proposition \ref{thm:normkotz}}. Proposition \ref{thm:normkotz} follows if the conditions (N1-N4) by Huber (1967) are fulfilled.  The notation of this proof will be chose to match Huber's paper. Let $\cal{M}^+$ denote the set of symmetric positive definite $d\times d$ matrices. For $A \in \cal M^+$, we define its norm $\|A\|$ as the spectral norm of $A$, that is $\lambda_1$, where $\lambda_1\geq \lambda_2\geq ...\geq\lambda_d$ are eigenvalues of $A$. Without loss of generality, assume $\bi \mu=\bi 0$.   It is clear that the Kotz estimator  $\psi(\bi x, M)$ in Huber's paper takes the form of 
$$\psi (\bi x, M) =(\bi x^TM^{-1}\bi x)^{-1/2}\bi x\bi x^T-M.$$
 Let 
 $ \lambda(M) = \E \psi(\bi X, M)$ so that the true parameter $I_d$ is defined as $\lambda(I_d) =\bi 0$. Define 
 $$ U(\bi x, M,\delta) = \sup_{\|M_1-M\|<\delta}\|\psi(\bi x,M_1) -\psi(\bi x, M)\|.$$
According to Huber's Theorem 3 and its corollary, if there exist positive number $b$, $c$ and $\delta_0$ such that $\E U(\bi X, M, \delta)<b\delta$ and $\E U^2(\bi X, M, \delta) < c\delta$ for $\|M-I_d\|+\delta<\delta_0$ and if 
$\E (\|\psi(\bi X, I_d)\|^2)$ is nonzero and finite,
then the asymptotic normality of $\hat{\Sigma}_K$ follows.

Note that $U(\bi x, M, \delta)$ is less than
$$\delta+\frac{\|\bi x\bi x^T\|}{\|\bi x\|} \sup_ {\|M_1-M\|<\delta} |(\frac{\bi x^TM_1^{-1}\bi x}{\bi x^T\bi x})^{-1/2}-(\frac{\bi x^TM^{-1}\bi x}{\bi x^T\bi x})^{-1/2}|.  $$
Hence, for sufficient small $\delta$, $\E U(\bi X, M, \delta) < \delta+\E \|\bi X\|/d  \max(\sqrt{\lambda_1+\delta}-\sqrt{\lambda_d}, \sqrt{\lambda_1}-\sqrt{\lambda_d-\delta} )$, where $\lambda_1$ and $\lambda_d$ are the largest and smallest eigenvalues of $M$, respectively. Since $\|M-I_d\|\leq \delta_0-\delta$, we have $\lambda_1<1+\delta_0-\delta$ and $\lambda_d>1-\delta_0+\delta$. Thus, $\E U(\bi X, M, \delta)< \delta+\E \|\bi X\|/d \max (\sqrt{1+\delta_0}-\sqrt{1-\delta_0+\delta}, \sqrt{1+\delta_0-\delta}-\sqrt{1-\delta_0})$, and it exists a $b>0$ such that $\E U(\bi X, M, \delta)<b\delta$.  Similarly, the existence of $c$ can be proved under the assumption of finite second moment, that $\E \|\bi X\|^2<\infty$. Also the result that $\E (\|\psi(\bi X, I_d)\|^2)$ is nonzero and finite follows for continuous $F_0$ with a finite second moment. 


\end{document}